\documentclass[twocolumn,twoside,slac_two]{revtex4}
\usepackage{graphicx}
\usepackage{natbib}
\usepackage{fancyhdr}
\newcommand{\sun}{\ensuremath{\odot}}%                          % sun symbol
\begin{document}

\title{A search for gravitational lensing effects in Fermi GRB data}


\author{P. Veres}
\affiliation{E\"otv\"os University, Budapest, Bolyai Military University, Budapest}
\author{Z. Bagoly}
\affiliation{E\"otv\"os University, Budapest}
\author{I. Horv\'ath}
\affiliation{Bolyai Military University, Budapest}
\author{A. M\'esz\'aros}
\affiliation{Charles University, Prague}
\author{L. G. Bal\'azs}
\affiliation{Konkoly Observatory, Budapest}
\begin{abstract}
      As GRBs trace the high-z Universe, there is a non-negligible probability of a lensing effect being imprinted on the lightcurves of the bursts.
      We propose to search for lensed candidates with a cross-correlation method, by looking at bursts days to years apart coming from the same part of the sky.
      We look for similarities and hypothesize a Singular Isothermal Sphere (SIS) model for the lens. 
      A lensed pair would enable us to constrain the mass of the lensing object.
      Our search did not reveal any gravitationally lensed events.
\end{abstract}

\maketitle

\thispagestyle{fancy}


\section{Lensing basics}
       The detected gravitationally lensed quasar images have a separation which is smaller than the resolution of current gamma-ray detectors.
       The time resolution of gamma-ray instruments however is superior in many aspects to the optical instruments.
       The time-delays we are looking for, come in two flavours that define two types of searches.
       We can compare the shape and spectra of two bursts coming from the same part of the sky or we can
       look for similarly shaped pulses in a single GRB's stream of photons. In this work we focus on the first scenario. We just mention the
       second case for its simplicity.  The equation from \cite{1991ApJ...378...22K} defines the characteristic mass scale of the point lens up to a factor $(1+z)$.
      \begin{equation}
        M(1+z_{\mathrm{LENS}}) = \frac{ \Delta \tau c^3}{2G} \left(\frac{1-f}{\sqrt{f}} + \ln{f}\right)^{-1} 
      \end{equation}
      $\Delta \tau$ is the time delay between the pulses, $f(<1)$ is the ratio of the peak counts. This model is deemed accurate for time
      delays of $\sim 10^{-5}$ to $\sim100$ s with lens masses ranging from $\sim M_{\sun}$ to $\sim 10^7 M_{\sun}$.
      For time delays of days to months we use the a single isothermal sphere model. In this case it is impossible to ascertain the
      lens mass, only a combination if the angular diameter distances (lens-source ($D_{ds}$), observer-lens ($D_d$)
      and observer-source ($D_s$)) and the velocity dispersion ($\sigma_v$):
      \begin{equation}
        \frac{\Delta \tau (1-f)}{2(1+f)} = \frac{ (1+z_{\mathrm{LENS}})}{c} \frac{D_d D_{ds}}{D_s} \left(\frac{4\pi \sigma_v^2}{c^2}\right)^2
      \end{equation}
	  In this work we will be searching for bursts from the same part of the sky.
\section{How to find a lensed burst?}
      We used the bursts in the Fermi/GBM burst table. We imposed an upper limit on the error radius of $6^\circ$.
      We selected pairs of bursts closer than two-times their positional errors ($2\sigma$).
      The next step was creating pairs of lightcurves and their cross-correlation
      curve on the same timescale for comparison. Each of the resulting $179$ figures was inspected for similarities
      with the aid of the cross-correlation curve at two timescales ($0.064$ and $0.512$ s). If the burst was deemed interesting, we proceeded
      to compare the spectral parameters by fitting for the spectrum. Simultaneously we check if the early burst is brighter than the later.
      This is a necessary condition for the time delays \citep{1992ApJ...389L..41M}.

         \begin{figure*}[t]
          \centering
          \includegraphics[width=135mm]{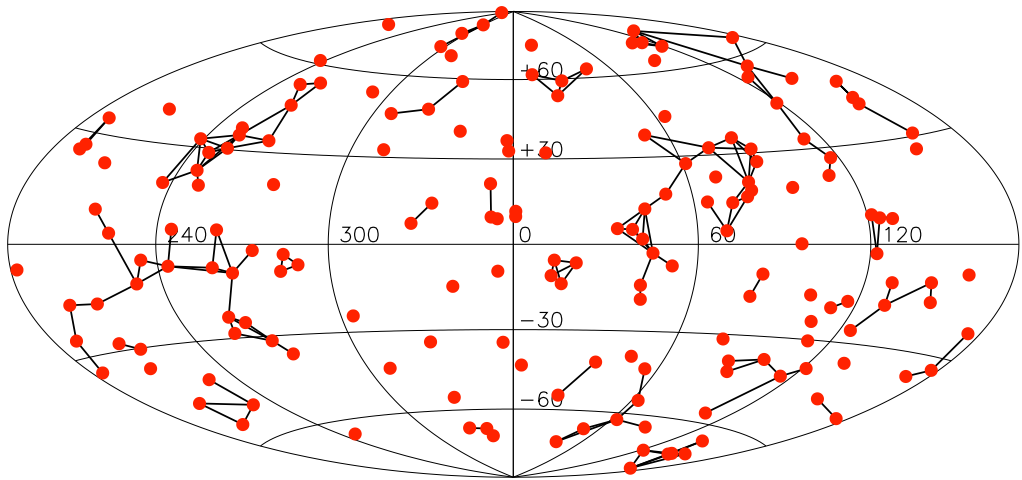}%\\ \bigskip\vspace{0.2cm}
          \caption{Figure showing all the Fermi GRBs, with the ones closer than 2 times their position errors connected.}
           \end{figure*}
       The typical echo timescale is from days to months but no longer than $\sim 14$ months (the elapsed time from the first Fermi trigger).
      \section{Candidate events}

\begin{center}
\begin{tabular}{|c|c|c|c|c|}
\hline
\multicolumn{2}{|c|}{Triggers}                  & Distance                  & $\Delta \tau [\mathrm{days}]$ &   f  \\ \hline \hline
080730.786  & 090730.608    &  $5.5^\circ (0.9\sigma)$  & $364.8$                       & $0.29$        \\
081216.531  & 090429.753    &  $4.8^\circ (0.5\sigma)$  & $134.2$                       & $0.44$        \\
090516.853  & 090514.006    &  $4.3^\circ (0.5\sigma)$  & $2.9$                         & $\lesssim 1$\\ \hline
      \end{tabular}\\
      %\caption
      \end{center}

\section{Discussion}

    As in \cite{1994ApJ...432..478N}, a list of close calls was established. Examples can be seen in the table below.
    None of these pairs passed the spectral inspection convincingly. We have relaxed our $2\sigma$ criterion to $3\sigma$ and checked the resulting
    $270$ pairs, but found no additional candidates. \\

	\subsection{080730.786 and 090730.608}
		 Both of these bursts have three peaks, with roughly the same time
        between them. It is curious that the elapsed time between them is roughly one year. We fitted a Band spectrum for the
        first two peaks. We compare the low-energy spectral indices($\alpha$) as both bursts have most of their emission in this range and not all of the other
        spectral parameters could be constrained. While for 080730 $\alpha$ decreases slightly from $-0.53\pm0.1$ to $-0.56\pm 0.1$, for 090730 $\alpha$ decreases
        (from $-0.74\pm0.2$ to $-0.63\pm 0.3$). Even though the indices have some overlap, we feel it is hard to put forward convincing arguments
        for a lensing scenario in this pair.\\
         \begin{figure}[ht]
          \centering
          \includegraphics[width=.9\columnwidth]{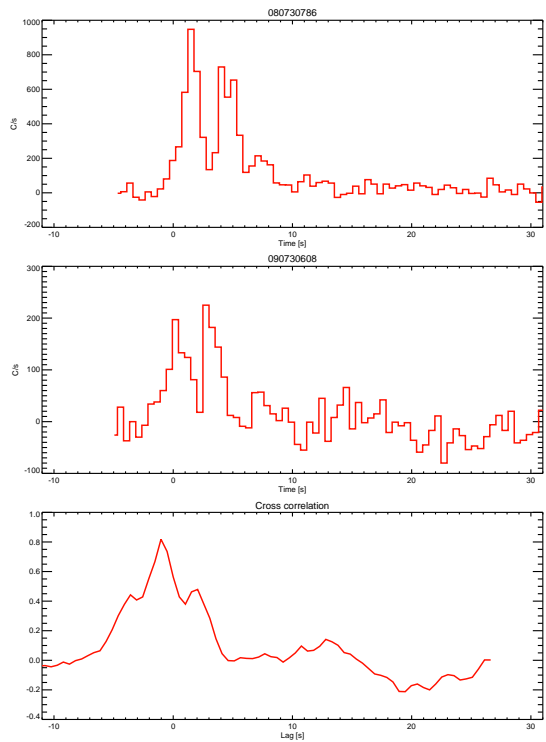}\\%\bigskip\vspace{0.2cm}
          \caption{Fermi GRBs 080730.786 and 090730.608 with their most luminous detector and their cross correlation function.The resolution is $0.512$ s.}
           \end{figure}

       \subsection{081216.531 and 090429.753}
		 These short bursts were also put forward as worthy for a spectral inspection. It turned out that 090429 has a
        significantly higher peak energy ($E_p=1235\pm264$ keV \textit{vs.} $E_p=152.8\pm92.4$ keV ), though the spectral indices agree fairly well
        ($\alpha_1=-0.66\pm0.53$, $\alpha_2=-0.70\pm0.09$ and $\beta_1=-1.88\pm0.32$, $\beta_2=-2.17\pm0.21$)
        \citep{2008GCN..8680....1M,2009GCN..9311....1B}\\
         \begin{figure}[ht]
          \centering
          \includegraphics[width=.9\columnwidth]{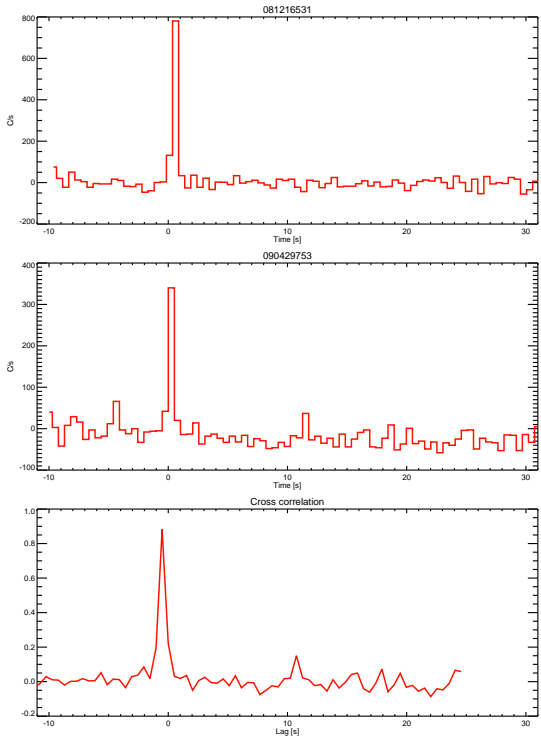}\\%\bigskip\vspace{0.2cm}
          \caption{Fermi GRBs 081216.531 and 090429.753 with their most luminous detector and their cross correlation function. The resolution is $0.512$ s.}
           \end{figure}

        \subsection{090516.853 and 090514.006}
		There is no detector response matrix (DRM) available for 090516.853 and no GCN was published for this trigger.
        This prevented us from spectral fitting and from determining the burst-detector angle, hence the uncertainty on $f$. In spite of a visible extended emission
        region unique to the lightcurve of 090514.006, we considered this a good candidate based on the similarities of the profile of the main event.
        We are waiting for the release of the DRMs complete the analysis.
         \begin{figure}[hb]
          \centering
          \includegraphics[width=.999\columnwidth]{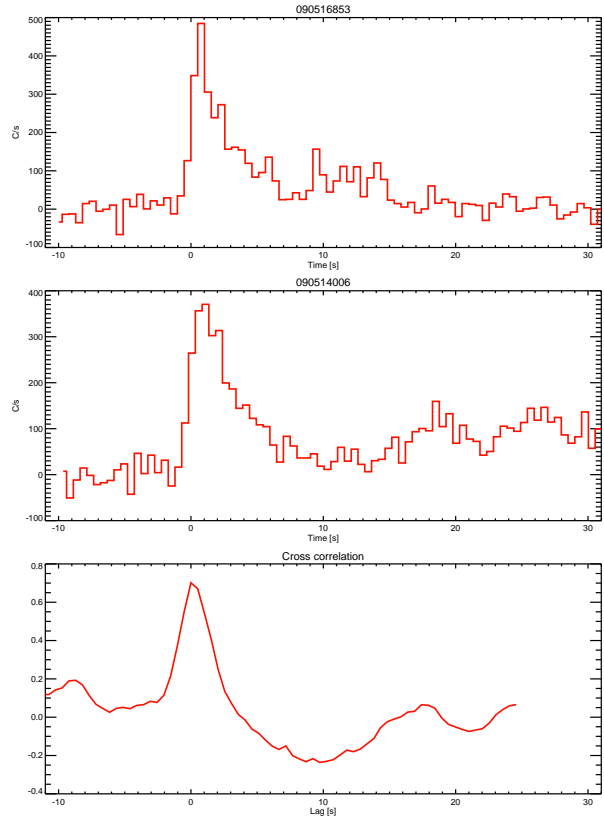}\\%\bigskip\vspace{0.2cm}
          \caption{Fermi GRBs 090516.853 and 090514.006 with their most luminous detector and their cross correlation function. The resolution is $0.512$ s.}
           \end{figure}

\section{Conclusion}
    We carried out a search for lensing signatures in an up-to-date burst sample. We selected bursts closer than $2 \sigma$ and found three presentable
    candidates for a lensing events that occurred at long timescales (days to months).
    Inspection of the spectra casted doubts on the viability of the lensing scenario in two cases and we need more data to decide in a third case.

  \begin{acknowledgments}
        This research is supported by Hungarian OTKA grant K077795, by the Bolyai Scholarship (I. H.), by the GAUK grant No. 46307, and by the
        Research Program MSM0021620860 of the Ministry of Education of the Czech Republic (A.M.).
  \end{acknowledgments}

%\bibliography{lensing}

% If in two-column mode, this environment will change to single-column
% format so that long equations can be displayed. Use
%\begin{widetext}
%\end{widetext}

\end{document}